\documentstyle[12pt,aasms4]{article}

\lefthead{Bekki Kenji}
\righthead{Origin of companion galaxies}

\begin{document}
\title{Origin of companion galaxies in QSO hosts}

\author{Kenji Bekki} 
\affil{Astronomical Institute, Tohoku University, Sendai, 980-8578, Japan} 


\begin{abstract}

Recent morphological studies of QSO host galaxies by the Hubble Space
Telescope (HST) have revealed that a sizable fraction of QSO host galaxies
possess close small companion galaxies. 
It is however not clear why and how these companion galaxies
are physically associated with the activation of QSO nucleus and the formation
of QSO hosts.
We here demonstrate  that QSO companion galaxies are 
tidally formed by a gas-rich major galaxy merger
and then orbit the merger remnant, 
based on the results of numerical simulations which
investigate both gas fueling  to the central 
seed massive black holes in  the  merger 
and global morphological evolution of the merger. 
We furthermore suggest that some of companion galaxies tidally formed
in a  major merger can evolve eventually  into small compact elliptical galaxies
observed frequently  in the present-day  bright massive  galaxies.
Thus our numerical simulations not only clarify the origin 
of companion galaxies observed in low and intermediate redshift
QSOs but also provide an evolutionary link between QSO companion galaxies
and the present-day compact elliptical galaxies located in the vicinity of  giant elliptical
galaxies.

\end{abstract}


\section{Introduction}
 Since faint nebulosity around quasars was discovered (Matthews \& Sandage 1963; Sandage \& Miller 1966),
 morphological studies of QSO host galaxies
 have revealed 
 the evolutionary
 link between the formation of QSO hosts and the activation  of QSO nucleus
 (Hutchings et al. 1982; 
 Malkan 1984; 
 Margon, Downes, \& Chanan 1984; 
 Smith et al. 1986;
 Heckman et al. 1991).
 Photometric and spectroscopic studies of QSO hosts furthermore 
 provided valuable clues to the nature of stellar populations of QSO host
 (MacKenty \& Stockton 1984; Boronson, Perrson, \& Oke 1985;
  Stockton \& Rigeway 1991; Dunlop et al. 1993; McLeod \& Rieke 1994).
 One of the most remarkable observational evidences is 
 that galaxy interaction and merging can trigger the nuclear activities
 of QSOs (Stockton 1982; Hutchings \& Campbell 1983; Stockton \& Mackenty1983;
 Hutchings \& Neff 1992; Bahcall et al. 1997).
 In particular, the recent  high-resolution morphological studies of
 QSO host galaxies by the Hubble Space Telescope (HST) 
 and large grand-based ones found 
 that a sizable fraction of QSO hosts have close  companion galaxies
 likely to be interacting or merging with the hosts (Bahcall et  al. 1995; Disney et al 1995).
 Although these  observational studies  strongly suggest that  close companion galaxies 
 in QSO hosts play a vital role in triggering QSO activities (Bahcall et  al. 1995),
 it is  still theoretically unclear why QSO host galaxies so frequently have
 companions and how QSO activities are physically associated
 with the formation and the evolution of such companion galaxies.


 In this Letter, 
 we numerically investigate both gas fueling to 
 the seed black holes located in the central part of two
 disks in a gas-rich merger and the morphological evolution
 of the merger in order to
 present a plausible interpretation on the origin
 of small companion galaxies frequently observed in QSO host galaxies.
 We here  demonstrate that the observed QSO companion galaxies are 
 formed in the outer part of strong tidal tails during gas-rich major galaxy merging
 and then become self-gravitating compact galaxies orbiting elliptical
 galaxies formed by merging. 
 We furthermore demonstrate that such companion galaxies are located
 within a few tens kpc of elliptical galaxies when efficient gas fueling
 to the central seed QSO  black holes continues.
 We thus suggest that both the formation of QSO companion galaxies
 and the activation of QSO nucleus 
 result from one physical process of gas-rich major galaxy merging.
 We furthermore  discuss whether such companion galaxies formed in QSO hosts
 can finally become compact elliptical galaxies that are  frequently
 observed in the present-day  bright massive galaxies.

\section{Model}

 We construct  models of galaxy mergers between gas-rich 
 disk galaxies with equal mass by using Fall-Efstathiou model (1980).
 The total mass and the size of a progenitor disk are $M_{\rm d}$
 and $R_{\rm d}$, respectively. 
 From now on, all the mass and length are measured in units of
  $M_{\rm d}$ and  $R_{\rm d}$, respectively, unless specified. 
  Velocity and time are 
  measured in units of $v$ = $ (GM_{\rm d}/R_{\rm d})^{1/2}$ and
  $t_{\rm dyn}$ = $(R_{\rm d}^{3}/GM_{\rm d})^{1/2}$, respectively,
  where $G$ is the gravitational constant and assumed to be 1.0
  in the present study. 
  If we adopt $M_{\rm d}$ = 6.0 $\times$ $10^{10}$ $ \rm M_{\odot}$ and
  $R_{\rm d}$ = 17.5 kpc as a fiducial value, then $v$ = 1.21 $\times$
  $10^{2}$ km/s  and  $t_{\rm dyn}$ = 1.41 $\times$ $10^{8}$ yr,
  respectively.
  In the present model, the rotation curve becomes nearly flat
  at  0.35  $R_{\rm d}$  with the maximum rotational velocity $v_{\rm m}$ = 1.8 in
  our units.
  The corresponding total mass $M_{\rm t}$ and halo mass $M_{\rm h}$
  are 5.0  and 4.0 in our units, respectively.
  The radial ($R$) and vertical ($Z$) density profile 
  of a  disk are  assumed to be
  proportional to $\exp (-R/R_{0}) $ with scale length $R_{0}$ = 0.2
  and to  ${\rm sech}^2 (Z/Z_{0})$ with scale length $Z_{0}$ = 0.04
  in our units,
  respectively.
  The Toomre's parameter (\cite{bt87}) for the initial disks is
  set to be 1.2.
  The collisional and dissipative nature 
  of the interstellar medium is  modeled by the sticky particle method
  (\cite{sch81}).
  Star formation 
  is modeled by converting  the collisional
  gas particles
  into  collisionless new stellar particles according to 
  the Schmidt law (Schmidt 1959)
  with the exponent of  2.0.
  The initial gas mass fraction ($f_{\rm g}$) is considered to be a free parameter ranging
  from 0.1 (corresponding to a gas poor disk) to  0.5 (a very gas-rich one).
  We here present the result of the model with $f_{\rm g}=0.5$, because 
  this model  most clearly shows  the typical behavior of QSO companion formation.
  The dependence of the details of QSO companion formation on $f_{\rm g}$ will be
  described by our future paper (Bekki 1999).

 The orbital plane of a   galaxy merger is assumed to be the same as $xy$ plane
 and the initial distance between the center of mass of merger progenitor
 disks is 8.0 in our units (140 kpc).
 Two disks in the merger are assumed to encounter each other parabolically 
 with the pericentric distance of 1.0 in  our units (17.5 kpc). 
 The intrinsic spin vector of one galaxy in a  merger  is exactly parallel with $z$ axis
 whereas that of  the other  is tilted by ${30}^{\circ}$ from $z$ axis.
 The present study describes  the QSO companion formation
 only for  a nearly prograde-retrograde merger in which only
 one intrinsic spin vector of a merger progenitor galaxy is nearly parallel with orbital
 spin vector of the merger.
 The dependence of the details of QSO companion formation processes on the initial
 orbital configurations of galaxy mergers will be given by Bekki (1999).
  The  number of particles used in a simulation 
  is  20000 for dark halo components, 20000 for stellar ones,
  and 20000 for gaseous ones.
All the calculations including the dissipative and dissipationless
dynamics and star formation
 have been carried out on the GRAPE board
  (\cite{sug90})
  at Astronomical Institute of Tohoku University.
  The parameter of gravitational softening is set to be fixed at 0.03
  in all the simulations. 

  By using this merger model, we firstly investigate    
  morphological and dynamical evolution of a gas-rich major
  galaxy merger  with a particular emphasis on the 
  formation of close small companions  (dwarf-like galaxy) in the merger.
  Secondly, we investigate when and how QSO activities are triggered
  by major galaxy merging by counting total mass of interstellar gas  accumulated within
  the central 100 pc of a galaxy merger.
  In order to estimate the gas mass in a explicitly self-consistent manner,
  we initially  place a collisionless particle with the mass equal to  $3.0 \times {10}^{6}$
  in the mass center of a disk and regard this particle as a `seed black hole'.
  We then investigate both the time evolution of the orbit of the seed black hole and the total gas mass 
  transferred to the central 100 pc around  the black hole.
 Here we hypothetically assume that interstellar  gas transferred to the central 100 pc around  the seed black hole 
 can be furthermore fueled to the central sub-pc region where a massive black hole
 gravitationally dominates and utilizes gas falling onto the accretion disk for a QSO activity. 
 The reason for our adopting this assumption is that we regard 
 a certain mechanism for gas fueling to the sub-pc region, such as the so-called
 `bars within bars'
 proposed by Shlosman, Frank, \& Begelman (1989), as being occurred  
 naturally  in the high-density self-gravitating central regions of galaxy mergers. 
 The above two-fold investigation just allows us to address questions as to
 when and how galaxy merging not only  forms  small  companions  but also  triggers  QSO nuclear activities.

\placefigure{fig-1}
\placefigure{fig-2}

\section{Result}
 Figure 1 describes how a QSO companion galaxy is formed by gas-rich major 
 galaxy merging. As two gas-rich disks merge to form a tidal tail composed
 of gas and stars (the time $T$ = 1.1 Gyr), the stellar components in the tail first collapse to form
 a self-gravitating dwarf-like object. Gaseous components are then swept
 into the deep gravitational potential well of the dwarf galaxy 
 to form a massive gaseous clump owing to the enhanced gaseous dissipation in
 the shocked region of the tidal tail and the dwarf.
 Star formation proceeds very efficiently in the 
 high density gas clump, 
 and consequently new stellar components  are  formed in the dwarf galaxy ($T$ = 1.7 Gyr). 
 The physical processes of  the dwarf galaxy formation in the present star-forming
 galaxy merger are essentially the same as those described by Barnes \& Hernquist
 (1992).
 This self-gravitating dwarf galaxy can then orbit an elliptical galaxy formed by galaxy merging
 without significant radial orbital decay due to dynamical friction
 between the dwarf  and the host elliptical and tidal destruction
 by the elliptical ($T$ = 2.3  and 2.8 Gyr).
 Total mass in the dwarf at $T$ = 2.3 Gyr is  roughly estimated to be  
 $\sim 2.7 \times {10}^{9} {\rm M}_{\odot}$ corresponding to 4.5 \% of the initial disk
 mass. The gas mass fraction of the dwarf is rather  large ($\sim 25\%$), which
 reflects the fact of the dwarf's being formed in the gas-rich tidal tail.
 About 45 \% of stellar components of the dwarf are very young stars formed from gaseous components
 of the tidal tail,
 which implies that 
 this dwarf galaxy 
 can be  observed to show very blue colors until its hot and massive stars died out.
 Considering that the present gas-rich star-forming merger model also shows  efficient
 gas fueling to the central seed black holes (as is described later),
 we regard the above  results as demonstrating  clearly that 
 the dwarf galaxy formed in galaxy merging can be observed as a companion galaxy 
 in a QSO host galaxy.

 Figure 2 shows the star formation history of the merger and the time evolution
 of gas mass located within  the central 100pc around the seed black holes
 of the merger.
 Star formation rate becomes maximum ($\sim 378 \rm M_{\odot}/\rm yr$) at $T$ = 1.3 Gyr,
 when two disks finally merge to form an elliptical galaxy
 and the efficient redistribution of angular momentum and gaseous  dissipation
 by cloud-cloud collisions cooperate  to form the  extremely high-density gaseous regions
 in the central part of the merger.
 After the intense secondary starburst, the star formation
 then rapidly declines owing to the efficient  gas consumption by the starburst.
 Gas fueling  to the central seed black holes  becomes maximum 
 ($6.5 \times {10}^{8} \rm M_{\odot}$) at $T$ = 1.3 Gyr,
 which is the same as  the maximum starburst of the merger.
 Gas supply for the seed black holes is greatly controlled by the rapid gas
 consumption by star formation,
 and consequently gas fueling gradually declines 
 after the completion of the secondary starburst.
 The  gas fueling in the present study
 tends to be  more efficient in the late phase of galaxy merging ($ T > 1.3 $ Gyr) 
 than in the early one ($ T < 1.3 $ Gyr).
 Assuming that all of the  gas  transferred to the central 100pc
 around  the seed black holes 
 can be directly accreted onto
 the accretion disk of the black holes,
 we can estimate that the mean accretion rate in the merger late phase (1.3 Gyr $<T<$ 2.3 Gyr) 
 is  $6.3 \rm M_{\odot}/\rm yr$.
 The derived accretion  rate is sufficient enough to trigger the typical magnitude of QSO activity 
 (e.g., Rees 1984).
 These  results imply that  secondary massive starburst and QSO nuclear activity (AGN) can be observed to coexist
 in a QSO host galaxy, which is consistent with the observational evidence
 that some of QSO host galaxies show very bluer colors and spectroscopic properties indicative
 of the past starburst (MacKenty \& Stockton 1984; Boronson, Perrson, \& Oke 1985;
 Stockton \& Rigeway 1991).

  Thus Figure 1 and 2 clearly demonstrate that gas-rich major galaxy merging
  not only contributes to the formation of a companion galaxy orbiting a merger remnant
  but also triggers QSO nuclear activities.
  Accordingly our  numerical study 
  can naturally explain why QSO host galaxies, some  of which are actually observed to be ongoing
  mergers and elliptical galaxies (e.g., Bahcall et al. 1997),
  are more likely to have close small companion galaxies;
  This is essentially because both QSO host galaxies with pronounced nuclear activities
  and their companions
  result from $one$ physical process of major galaxy merging.
  Our numerical studies furthermore provide the following three  predictions 
  on physical properties of QSO companions and hosts.
  First prediction is that the  luminosity of a QSO companion galaxy
  is roughly proportional to that of the QSO host,
  principally because the mass of tidal debris that is a progenitor of a QSO companion
  depends strongly on the initial mass of a galaxy merger.
  Second is that a QSO companion has very young stellar population formed in secondary
  starburst of galaxy merging and thus shows photometric and spectroscopic properties
  indicative of starburst or post-starburst.
  Third is that not all of galaxy mergers can create QSO companions galaxies,
  essentially because nearly retrograde-retrograde mergers can not produce
  strong tidal tails indispensable for the formation of companion galaxies
  because of the weaker tidal perturbation of the mergers (The details of
  the physical conditions required for the formation of QSO companions
  will be  described in  Bekki (1999)).
 We suggest that future observational studies on the dependence of the luminosity-ratio
 of QSO hosts to QSO companions on QSO host luminosity,
 age and metallicity distribution of stellar populations of QSO companions,
 and the probability that QSO host galaxies have companion galaxies physically associated
 with them can
 verify  the above three  predictions and thereby can determine whether 
 major galaxy merging is a really plausible model of  QSO companion formation.

\section{Discussion and Conclusion}

 The fate of QSO companion galaxies is an interesting problem of the present
 merger scenario of QSO companion formation.
 We here propose that some of the companions finally evolve into compact
 elliptical galaxies (cE) that have  typical blue magnitude
 $M_{\rm B}$ ranging from -18 mag to -14 mag,  truncated de Vaucouleurs luminosity profile,
 color-magnitude relation of giant ellipticals, typically solar-metallicity,
 and higher degree of global rotation (Faber 1973; Wirth \& Gallagher 1984;
 Nieto \& Prugniel 1987; Freedman 1989; Bender \& Nieto 1990;
 Burkert 1994).
 The essential reason of this proposal is described as follows.
 Burkert (1994) numerically investigated the dynamical evolution of proto-galaxies
 experiencing  an initial strong starburst
 and the  subsequent violent relaxation in the tidal external gravitational field
 of a massive elliptical galaxy
 and revealed that the observed peculiar properties of cEs are due to the external
 tidal field around progenitor proto-galaxies  of cEs.
 Burkert (1994)  accordingly proposed a scenario in which satellite
 proto-galaxies revolving initially around a bright elliptical
 galaxy eventually form cEs after violent cold collapse and strong starburst around the galaxy.
 Although his model of cE formation is  not directly related to physical processes
 of gas-rich major galaxy merging,
 the physical environment of cE formation in his model is very similar to
 that of gas-rich galaxy  merging; Tidal debris collapses to form a self-gravitating small galaxy
 in the rapidly changing external gravitational field of two merging disk galaxies
 in the present study.
 Accordingly it is not unreasonable to consider that some of  companion galaxies created 
 in tidal tails finally become cEs orbiting  elliptical galaxies  formed by major
 galaxy merging.
 The observational fact that   cEs exist almost exclusively as satellites  of bright
 massive galaxies 
 (Faber 1973;  Burkert 1994) strengthens the validity of the proposed evolutionary link between
 QSO companions and cEs.
 Furthermore, 
 the larger degree of global rotation in kinematics  observed in cEs (e.g., Bender \& Nieto 1990)  
 seems  to be consistent with the proposed scenario, since QSO companions
 are created in the tidal debris of  rotationally supported disk galaxies in the scenario.
 The present numerical study unfortunately cannot investigate in detail structural and kinematical
 properties of companion galaxies formed in galaxy mergers because of very small particle
 number of the simulated companion  ($\sim 800$).
 Our future high resolution simulations with the total particle number of $\sim {10}^{7}$
 will enable us to compare the numerical results of structural,  kinematical, and chemical
 properties of QSO companions formed in major mergers with observational
 ones of cEs located near giant ellipticals  in an explicitly self-consistent manner
 and thereby answer  the question as to 
 the evolutionary link between intermediate and high redshift 
 QSO companions and the present-day cEs.

 The most important observational test to  assess
 the validity of the proposed formation scenario of QSO companion galaxies
 is to investigate whether a QSO companion galaxy 
 has  younger stellar populations formed by secondary starburst 
 and thus shows  photometric and spectroscopic properties indicative of starburst or post-starburst. 
 Canalizo \& Stockton (1997) 
 recently investigated  spectroscopic properties 
 of  companion galaxies in three QSOs (3CR 323.1, PG 1700+518, PKS 2135-147)
 and found that the spectra of a  companion galaxy in QSO PG 1700+518 
 shows both strong Balmer absorption lines from a relatively young stellar
 population and Mg I $b$ absorption feature
 and  the 4000 $ \rm \AA $ 
  break from an old stelar population.  
 Stockton, Canalizo, \& Close (1998) 
 furthermore demonstrated that the time that has elapsed since the end of the
 most recent major starburst event in the companion of QSO PG 1700+518
 is roughly 0.085 Gyr, based
 on the spectral energy distribution derived from adaptive-optics image
 in $J$ and $H$ band.
 These observational results on the post-starburst signature of QSO companions  are  
 consistent reasonably well with the proposed scenario 
 which predicts  that a QSO companion galaxy contains both relatively
 old stellar populations previously located in merger progenitor disks
 and very younger stellar populations formed in gas-rich  tidal tails. 
Detailed spectroscopic  studies of QSO companion galaxies, such as Canalizo \& Stockton (1997)
and Stockton, Canalizo, \& Close (1998),
have not been yet so accumulated.
Future  extensive  spectroscopic studies of companions in each of intermediate and high
redshift QSOs
will clarify the age distribution of stellar populations
of the companions and thus determine whether most of QSO  companions
are really formed in major galaxy mergers.

 We conclude that gas-rich major galaxy merging can naturally explain the prevalence
 of small companion galaxies in QSO hosts; The essential reason for the origin of QSO companions
 is that strong tidal
 gravitational field of major galaxy merging both triggers the formation of companions
 and provides efficient fuel for QSO nuclear activities.
 This explanation of QSO companion formation is consistent reasonably well with
 the observational fact that QSO nucleus are already activated though the companions
 are still located in the vicinity  of the QSO hosts ($\sim$ a few tens kpc from the center of the hosts).
 Our numerical simulations accordingly suggest that the observed companion galaxies in QSO
 hosts are not the direct $cause$ of QSO nuclear activities but the $result$ of gas-rich major galaxy merging.
 Although minor galaxy merging between small companion galaxies and giant elliptical galaxies or disk ones
 is demonstrated to be closely associated with secondary massive starburst in disks (Mihos \& Hernquist 1995)
 and strong starburst in shell galaxies (Hernquist \& Weil 1992), the present study
 implies that this minor merging is probably
 less important in the activation  of QSO nucleus  and the formation  of QSO companions.
 The present study provides  only one scenario of QSO companion formation,
 thus we lastly  stress that physical processes related to  
 the companion formation are  likely to be more variously different  and
 complicated than is described in the present study.

\acknowledgments

K.B. thank to the Japan Society for Promotion of Science (JSPS) 
Research Fellowships for Young Scientist.

\clearpage


\figcaption{
The morphological evolution of a major  galaxy merger 
between two gas-rich, bulgeless spirals with star formation
projected onto the $x$-$y$ plane.
In order to show more clearly the morphological evolution
of disk components, we do not display the halo components here.
Each frame measures 236 kpc (the scale is measured in units of 17.5 kpc),  and the time, indicated in the
upper left-hand corner of each panel is in units of gigayears.
Note that  the simulated morphology   
at $T=1.1$ Gyr and 
that at $T=2.8$ Gyr are remarkably  similar to the QSO 1403+434 (Hutchings \& Morris 1995) 
which is observed to have a dwarf-like  companion galaxy $within$  a tidal tail
associated with the QSO  
and to the 
intermediate redshift QSO PKS2128-123 (Disney et al. 1995)
which is observed to have a close compact companion
around the well relaxed elliptical host galaxy, respectively.
\label{fig-1}}

\figcaption{
The time evolution of  the star formation rate (upper panel)
and that of the total gas mass transferred to  the central 100 pc of
the seed black holes in a merger (lower panel).
Note that both the star formation rate and the central gas mass become
maximum at $T=1.3$ Gyr.
\label{fig-2}}

\end{document}